\documentclass[a4paper, 10pt, conference]{ieeeconf}      

\IEEEoverridecommandlockouts                              

\usepackage{url}
\usepackage{subfig}
\usepackage{graphics} 
\usepackage{epsfig} 
\usepackage{mathptmx} 
\usepackage{times} 
\usepackage{amsmath} 
\usepackage{amssymb}  
\usepackage{xcolor}
\usepackage{fixme}
\fxsetup{
    status=final, 
    author=,      
    theme=color,
}

\title{\LARGE \bf
MOSS: A Large-scale Open Microscopic Traffic Simulation System
}

\author{Jun Zhang$^{1}$, Wenxuan Ao$^{1}$, Junbo Yan$^{1}$, Can Rong$^{1}$, Depeng Jin$^{1}$, Wei Wu$^{2}$ and Yong Li$^{1\dagger}$
\thanks{$^{1}$Jun Zhang, Wenxuan Ao, Junbo Yan, Can Rong, Depeng Jin and Yong Li are with the Department of Electronic Engineering, Tsinghua University, Beijing 100084, China. (e-mail: 
\tt\small zhangjun990222@gmail.com, aowx21@outlook.com, yanjb20thu@gmail.com, rc20@mails.tsinghua.edu.cn, jindp@tsinghua.edu.cn, liyong07@tsinghua.edu.cn)}
\thanks{$^{2}$Wei Wu is with SenseAuto Research, Shanghai, China. (email: \tt\small wuwei@senseauto.com)}%
\thanks{$^{\dagger}$Yong Li is the Corresponding Author.}
}

\begin{document}

\maketitle
\thispagestyle{empty}
\pagestyle{empty}

\begin{abstract}

In the research of Intelligent Transportation Systems (ITS), traffic simulation is a key procedure for the evaluation of new methods and optimization of strategies.
However, existing traffic simulation systems face two challenges.
First, how to balance simulation scale with realism is a dilemma.
Second, it is hard to simulate realistic results, which requires realistic travel demand data and simulator.
These problems limit computer-aided optimization of traffic management strategies for large-scale road networks and reduce the usability of traffic simulations in areas where real-world travel demand data are lacking.
To address these problems, we design and implement \textbf{MO}bility \textbf{S}imulation \textbf{S}ystem (MOSS).
MOSS adopts GPU acceleration to significantly improve the efficiency and scale of microscopic traffic simulation, which enables realistic and fast simulations for large-scale road networks.
It provides realistic travel Origin-Destination (OD) matrices generation through a pre-trained generative neural network model based on publicly available data on a global scale, such as satellite imagery, to help researchers build meaningful travel demand data.
It also provides a complete open toolchain to help users with road network construction, demand generation, simulation, and result analysis.
The whole toolchain including the simulator can be accessed at \url{https://moss.fiblab.net} and the codes are open-source for community collaboration.

\end{abstract}

\section{Introduction}\label{sec:intro}

Traffic simulation is a critical foundational software for Intelligent Transportation Systems (ITS).
Researchers usually first use simulators to validate the advancement of their proposed methods or to optimize their strategies.
For instance, researchers in the field of transportation management improve signal control strategies by combining microscopic traffic simulation with reinforcement learning~\cite{van2016coordinated,wei2019colight,yazdani2023intelligent}.
Researchers in the field of CAVs evaluate their control algorithms for autonomous vehicles in simulation environments~\cite{tao2023flowsim,kusari2022enhancing,gulino2024waymax},
and also verify the vehicle collaboration strategies by it~\cite{cong2022virtual,zhang2022cavsim}.
In the existing simulation-based works, SUMO~\cite{behrisch2011sumo} is the most popular microscopic traffic simulator.
It is due to SUMO providing a complete pipeline necessary for microscopic traffic simulation.
Using SUMO, researchers can conveniently accomplish the four fundamental stages of traffic simulation: road network construction, demand generation, simulation, and result analysis.

However, when researchers extend the interested scale from a few intersections to large-scale road networks and use computer-aided optimization instead of manual design, existing traffic simulations face two important challenges:
\begin{itemize}
    \item \textbf{Difficulty in balancing simulation scale and realism.}
    Microscopic traffic simulation, such as SUMO and CityFlow~\cite{zhang2019cityflow}, employs car-following and lane-changing models to calculate the motion of each vehicle.
    This approach produces the most realistic results, but also requires the most extensive calculations.
    Meso-macroscopic traffic simulation, such as MATSim~\cite{w2016multi} and DynaSMART~\cite{mahmassani1992dynamic}, uses different levels of simplified models instead of vehicle-oriented models, reducing the calculation to achieve improved computational efficiency at the cost of reduced realism.
    Facing large-scale simulations that support computer-aided optimization tasks such as fine-grained traffic management, the existing methods are inadequate.
    The former is difficult to simulate in a finite amount of time and will make the optimization process extremely long for methods such as reinforcement learning.
    The latter will not be able to provide realistic microscopic results for fine-grained scenario optimization like traffic signal control.
    Therefore, it is important and urgent to implement an efficient and realistic traffic simulator.
    \item \textbf{Difficulty in simulating realistic result for large-scale road networks.}
    Simulating realistic results depends on realistic demand data and a realistic simulator.
    Currently, generating large-scale realistic demand data typically requires extensive traffic surveys or data acquisition through mobile phone network operators~\cite{iqbal2014development}, location-based service (LBS) providers~\cite{shi2020predicting}, or traffic cameras~\cite{yu2023city}.
    This prevents traffic simulation studies from being conducted in areas with underdeveloped digital infrastructures.
    Furthermore, even with qualified data, the realism of the simulator still determines to a large extent the realism and reliability of the simulation results.
    Therefore, providing both a realistic simulator and a globally available large-scale travel demand generation approach will significantly improve the usability of traffic simulation.
\end{itemize}


To solve the two challenges, we design and implement \textbf{MO}bility \textbf{S}imulation \textbf{S}ystem (MOSS) with the following main contributions:
\begin{itemize}
    \item We adopt GPU acceleration to implement large-scale microscopic traffic simulation with car-following and lane-changing models, significantly improves computational efficiency with guaranteed realism.
    From the experimental results, in the large-scale road network scenario with more than 2 million vehicles, the simulation speed of MOSS is 100.97 times higher than that of CityFlow, and 10.49 times higher than that of MATSIM.
    The advancement makes the model simplification previously used in meso-macroscopic traffic simulation for large-scale road networks unnecessary, thus solving the dilemma of simulation scale and realism.
    
    \item We address the worldwide Origin-Destination (OD) matrix generation problem in the demand generation pipeline through generative artificial intelligence (AI).
    Once the generative model has trained, it can generate realistic OD matrices for any region of interest around the globe based on publicly available data like satellite imagery.
    Based on the results, researchers can customize the traffic mode choice and route assignment to get meaningful travel demand data.
    According to the experimental results, the method improves the CPC and RMSE metrics by 20.50\% and 35.04\% compared to the best baseline model, respectively.
    Moreover, the OD matrices generated through it are highly correlated with ancillary data reflecting OD trips in several cities around the world.
    These results illustrate the strong realism and usability of the method, which can improve the realism of large-scale simulations.
    
    \item To help researchers make better use of MOSS and to make it open, we have open-sourced MOSS and the complete toolchain, including road network construction, demand generation, simulation and results analysis.
    To facilitate the construction of computer-aided optimization programs on top of MOSS, we have wrapped MOSS as a Python package and provided an large-scale transportation optimization example.
    We have also built an online wizard to help people try out the large-scale microscopic traffic simulation without coding.
\end{itemize}

\section{Related Works}\label{sec:related}

\subsection{Traffic Simulators}

Traffic simulation is an essential procedure for researchers to evaluate traffic system design, traffic management strategies, and new technologies for ITS.
Due to the large variations in the scale of road networks and the time span involved, existing traffic simulators can be categorized into three types based on the level of simplification of the simulation models used:
\begin{itemize}
    \item \textbf{Microscopic Simulators.}
    Microscopic simulators typically uses an interval of one second or less for simulation, independently computing each traffic participant, simulating behaviors such as car-following, lane-changing, acceleration, and deceleration. 
    The most popular open-source microscopic simulator is SUMO~\cite{behrisch2011sumo}, which has been used by researchers for decades in the field of ITS~\cite{van2016coordinated,tao2023flowsim,cong2022virtual}.
    To improve the computational efficiency of SUMO, CityFlow~\cite{zhang2019cityflow} implements simulation acceleration through multi-threaded parallelism, focusing on the traffic signal control optimization.
    Similarly, CBLab~\cite{liang2023cblab} further improves the simulation efficiency based on CityFlow.
    However, the computational efficiency of existing microscopic simulators is still limited by CPU computing power.
    When facing a large-scale scenario with about 1 million vehicles simulated at the same time, their single-step simulation will take close to wall-clock time.
    This makes simulation-based optimization impossible.

    \item \textbf{Mesoscopic Simulators.}
    Mesoscopic simulations usually simplify the vehicle motion model to speed up the simulation process.
    MATSim~\cite{w2016multi} is a widely-used mesoscopic simulator, using a uniform motion model and intersection waiting queues~\cite{gawron1998iterative} to model the movement of vehicles, omitting the details of acceleration and deceleration, which significantly improves the computational efficiency by about 10 times compared to multi-threaded microscopic simulators.
    However, such methods reduce the realism of the results to some extent, thereby affecting the reliability of the results of strategy optimization or method evaluation constructed on the basis.
    \item \textbf{Macroscopic Simulators.}
    Macroscopic simulators further simplifies the traffic system model by no longer considering the movement of individual vehicles, instead treating traffic as a fluid and describing it through concepts such as velocity and density.
    Popular macroscopic simulators include DynaSMART~\cite{mahmassani1992dynamic} and PTV VISUM\footnote{\url{https://www.ptvgroup.com/en/products/ptv-visum}}.
    Since vehicle modeling is discarded, the macroscopic simulators can only give statistical information about the road without vehicle motion data.
    This makes them unsuitable for applications in ITS that require fine-grained modeling, such as traffic management optimization at the intersection level.
\end{itemize}
In summary, existing methods fail to achieve large-scale traffic simulation.

\begin{figure*}[t]
  \centering
  \includegraphics[width=1\linewidth]{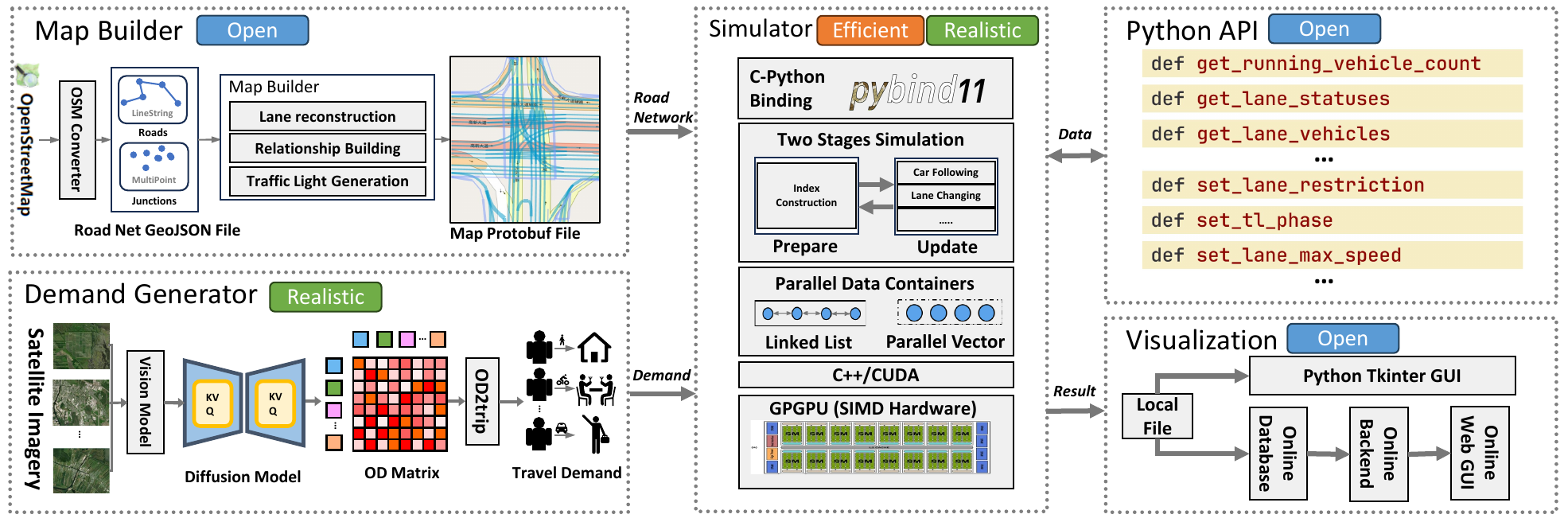}
  \caption{
  The framework and pipeline of MOSS.
  }\label{fig:moss}
  \vspace{-0.5cm}
\end{figure*}

\subsection{Origin-Destination Generation}\label{sec:related:demand}
Travel demand is an important input to traffic simulation.
Currently, one commonly applied method is the classical four-step demand generation, including trip generation, trip distribution, traffic mode choice and route assignment. 
The OD matrix is the result of the first two steps of the four-stage approach, which directly affects the realism of the demand generation results.
OD matrix generation approaches can be broadly categorized into two categories:
\begin{itemize}
    \item \textbf{Rule-based models}. 
    Existing classical rule-based models include gravity model~\cite{zipf1946p}, radiation model~\cite{simini2012universal}, and intervening opportunity model~\cite{stouffer1940intervening}.
    They all attribute population flows to some physical process and are therefore well interpretable.
    Rule-based models oversimplify the intricate population flow processes, making them inadequate for accurately representing complex population movements.  
\end{itemize}
\begin{itemize}
    \item \textbf{Data-driven models}.
    To improve the realism of generated results, data-driven models typically employ deep learning and neural network frameworks to uncover the inherent relationships between the real-world OD matrix data and input features.
    Features that are inputs to the model typically include social demographic statistics, geographical characteristics of each region in the target area, etc.
    Examples of such approaches include multi-graph convolution network~\cite{geng2019spatiotemporal}, random forest~\cite{pourebrahim2019trip}, deep neural networks~\cite{simini2021deep}, generative adversarial network~\cite{bojchevski2018netgan}, graph attention networks~\cite{liu2020learning} and diffusion model~\cite{rong2023city}. 
    Data-driven models often face limitations in training data and input feature acquisition, especially in developing countries and cities. 
\end{itemize}
Overall, the existing methods are not well qualified for generating realistic OD matrices for any region of interest around the globe due to data acquisition.
More accessible input data and more powerful and appropriate data-driven models adapted to such data may be viable directions to resolve the problem.

\section{MOSS: MObility Simulation System}\label{sec:moss}


In order to achieve efficient and realistic large-scale microscopic traffic simulation, we need to introduce a large amount of computational power far beyond the CPU to accelerate the calculation process of vehicle in order to solve the dilemma of scale and realism.
In addition, the travel demand generation process should be performed with only publicly available data as input to avoid the data lack problem, and the realism must be guaranteed.
Finally, the openness of the system and the completeness of the toolchain will determine whether users can really use the system.
We propose MObility Simulation System (MOSS) as shown in Figure~\ref{fig:moss} to achieve all the above goals.
First, MOSS adopts graphics processing units (GPUs) as the main computational hardware instead of CPUs, on which all simulation computation processes are highly parallelized to significantly accelerate the microscopic traffic simulation process with no cost of realism.
Second, MOSS adopts a highly transportable diffusion model combined with a large multi-modal model to realize the realistic OD matrix generation in any area of the globe based on publicly available data represented by satellite images.
Finally, MOSS builds a complete tool chain including road network construction, travel demand generation, simulation, visualization, Python package wrapped from the simulator, etc., which facilitates the users.
The design details will be introduced in the section.

\subsection{GPU-accelerated Simulator}


With the rapid development of general-purpose computing on graphics processing units (GPGPUs), parallel computing methods represented by single instruction and multiple data (SIMD) have become the mainstream approach for programmers to obtain huge amount computing power.
In the computational paradigm of microscopic traffic simulation, each vehicle is considered as an independent entity for computation, which is conducive to parallel computing.
Inspired by this, we believe that the computational acceleration of traffic simulation should shift from CPU-based multithreading solutions to GPU-based approaches due to the massive computational power that GPUs have.

Analyzing the computational characteristics of microscopic traffic simulation, we divide a computational iteration process in it into two parallel-friendly stages to avoid read-write conflicts and mutual exclusion protection in the vehicle sensing and decision making process.
The first stage, which is named the prepare phase, traverse all lanes to construct a linked list index structure based on the spatial relationships among vehicles to accelerate the process of vehicle sensing for neighboring vehicles in the update phase.
At the same time, a read-only copy of the vehicle's motion data is created to avoid read-write conflicts.
The second stage, which is named the update phase, execute the motion control procedures for all vehicles.
In this stage, vehicles use the index structure for rapid environment sensing, such as obtaining the speed and position of the preceding vehicle.
And then they calculate acceleration based on the car-following and lane-changing models.
Finally, they update their current position and velocity through Newton's equations of motion.
The car-following model we use is the IDM model~\cite{treiber2000congested}, and the lane-changing model is a randomized improvement of the MOBIL model~\cite{kesting2007general,feng2021intelligent}.
More technical details about parallel computing design can be referenced from our previous workshop paper~\cite{zhang2023city}.

The simulator requires input of road network data and demand data in Protobuf format.
It will simulate to obtain the velocity, position, and angle of all traffic participants at each moment (the default interval of moments is 1 second) as microscopic traffic information and also calculate macroscopic traffic measurements like the average speed of roads.

In terms of software implementation, the simulator is written in C++/CUDA and provides Python language bindings as a user-friendly programming approach through pybind11\footnote{\url{https://github.com/pybind/pybind11}}.

In short, leveraging the substantial computational power provided by GPUs, the simulator can significantly accelerate the process of microscopic traffic simulation, allowing researchers to no longer endure the slow simulation efficiency of large-scale scenarios or the necessity to compromise on realism through model simplification.

\subsection{OD Generator}


To overcome the problem of lacking data associated with existing OD generation methods that require high quality structured input features like sociodemographic and geographical data, MOSS's OD Generator adopts the cutting edge of generative AI to generate OD matrices from publicly available data.
In detail, on the one hand it adopts a cutting-edge and highly transferable diffusion model~\cite{rong2023city} to capture the intrinsic patterns of human travel activities worldwide.
On the other hand, it uses publicly available data represented by satellite imagery as a substitute for hard-to-access sociodemographic and geographical features as model input.
Utilizing recent advances in large multi-modal models for semantic image feature extraction~\cite{he2018perceiving,zhang2024uv}, satellite imagery possesses a highly powerful capability for urban area analysis and spatial characteristics acquisition, thereby providing input data for generative models on a global scale.
Consistent with existing work, the output of our proposed generative AI model remains the OD matrix.
For model training, we leverage satellite imagery as input and OD matrix data as ground truth, both obtained from within the United States.
As the sole input for the method, satellite imagery can be conveniently obtained from publicly accessible online platforms like Esri World Imagery\footnote{\url{https://www.arcgis.com/home/item.html?id=10df2279f9684e4a9f6a7f08febac2a9}}, Google Earth\footnote{\url{https://earth.google.com/}}, etc.
We choose Esri World Imagery as the data source of satellite imagery in MOSS.
The ground truth OD matrix data is collected and provided by the National Census Bureau~\cite{lodes2024us}, which will be described in detail in Section~\ref{sec:exp:demand}.
After completing the training, the model can output the realistic OD matrix by simply inputting data such as satellite images of the corresponding area.

Additionally, apart from the OD generation method based on generative AI and satellite imagery, we also provide other baseline models for users to choose from, including the gravity model, the radiation model, etc.

\subsection{Toolchain}

\subsubsection{Map Builder}


In order to easily adapt to common vector map formats on the one hand, and to make the final road network easy to be analyzed by computers on the other hand, a two-level road network data structure has been designed in MOSS.
The first level uses the GeoJSON standard\footnote{\url{https://geojson.org/}}, which facilitates convenient manual annotation and can be readily converted from publicly available data sources like OpenStreetMap (OSM)\footnote{\url{https://www.openstreetmap.org/}}.
The GeoJSON format describes the spatial location of the roads with information such as the number of lanes, speed limits, and describes the entering and exiting roads associated with each intersection.
We develop a tool to convert GeoJSON from OSM, which is capable to identify and create roads and junctions.
A visual editor is also developed for manual annotation\footnote{\url{https://moss.fiblab.net/tools/geojson-editor}}.

The second level is represented in Protobuf format\footnote{\url{https://protobuf.dev/}}, a compact binary format suitable for computer usage.
The Protobuf format extends the GeoJSON format with spatial locations and topological relationships for each lane and signal phases for intersections.
We develop a map builder to transform GeoJSON into Protobuf format, which can reconstruct lane connectivity within the intersection and generate traffic light signals. 
Leveraging OSM data, our tool is capable to convert OSM data to GeoJSON format, then generate with the map builder to road networks for most city in the world.

\subsubsection{Converter from OD matrix to travel demand}

To obtain individual travel demands suitable for simulation, we follow the four-step method and provide a customizable converter to do traffic mode choice, route assignment, etc., thereby converting OD matrices into individual travel demands.
The tool offers users the flexibility to adjust numerous parameters, including departure time distributions and proportions of transportation modes, etc.

\subsubsection{Visualization}

\begin{figure}[t]
	\centering
	\subfloat[Local visualization GUI]{\includegraphics[width=0.47\linewidth,height=4cm]{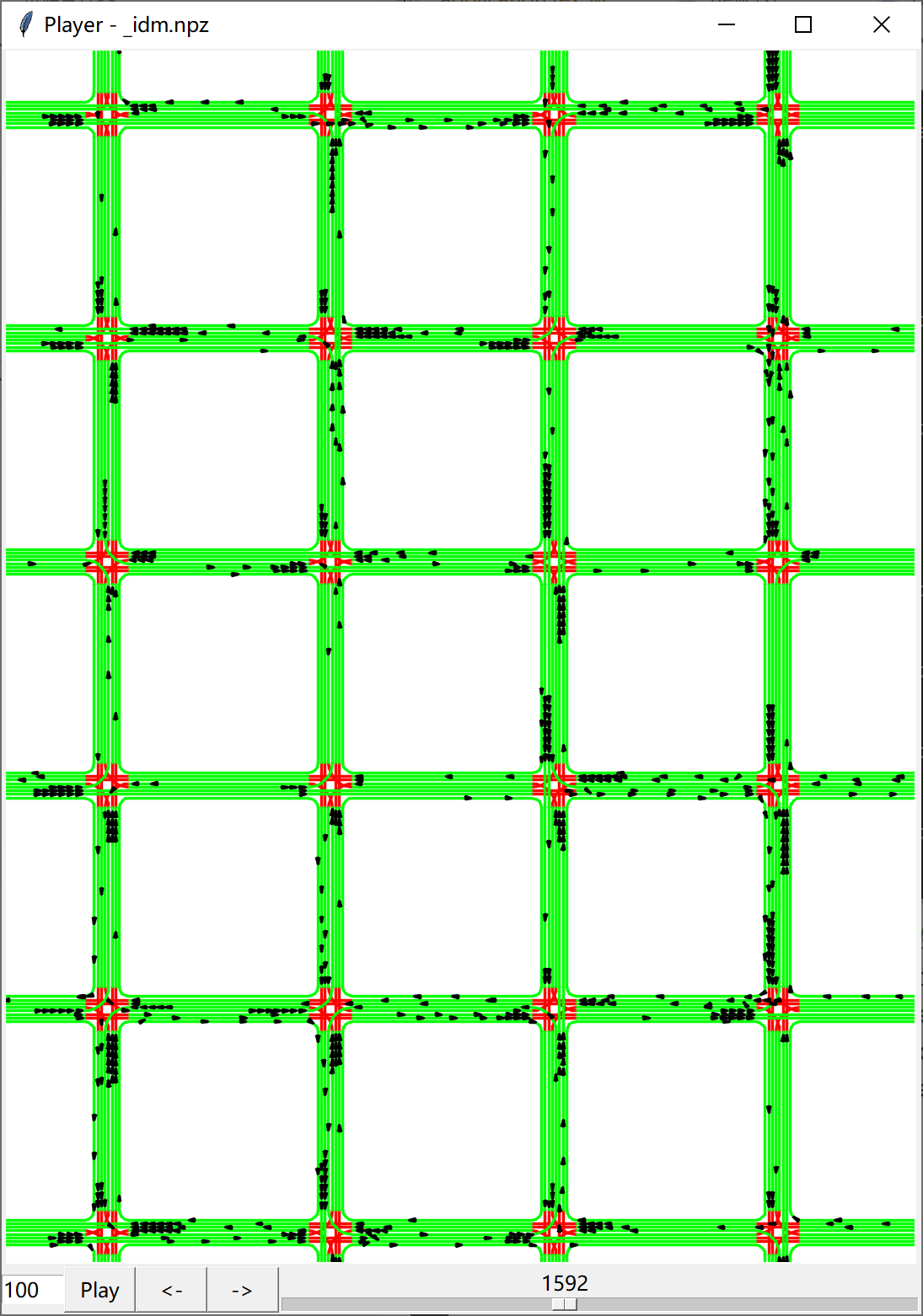}}
    \hspace{0.2cm}
	\subfloat[Online visualization GUI]{\includegraphics[width=0.47\linewidth,height=4cm]{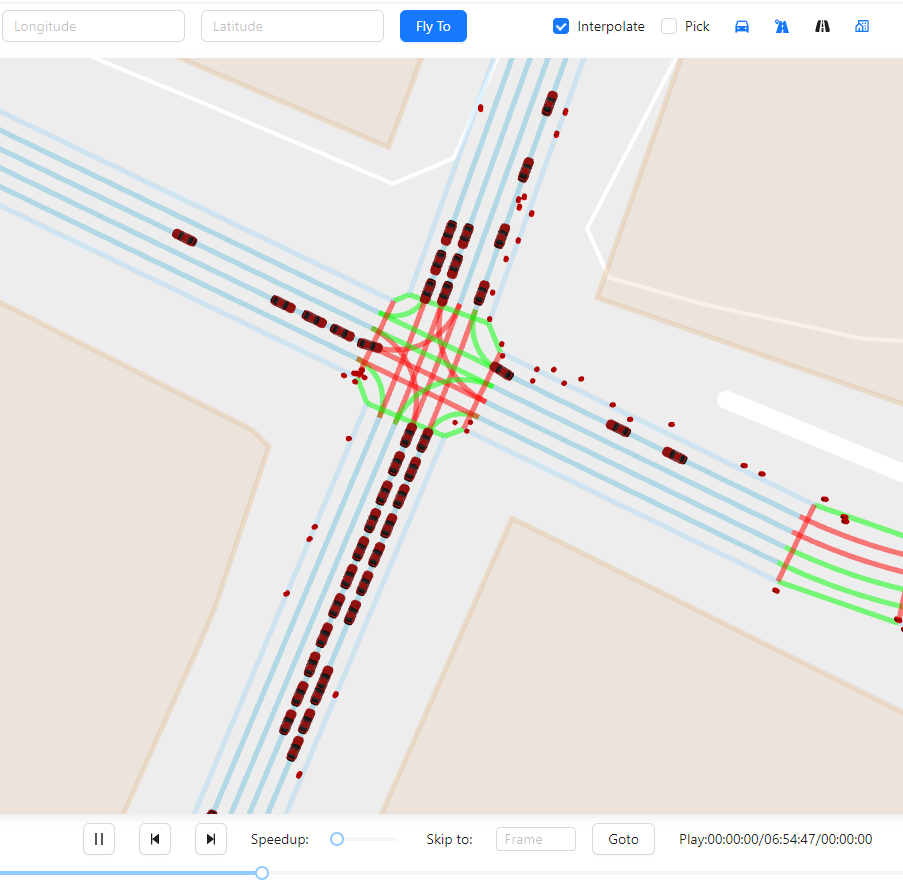}}
	\caption{Visualization GUIs provided by MOSS.}\label{fig:vis}
    \vspace{-0.5cm}
\end{figure}

For facilitating users to intuitively understand simulation results and verify their rationality, visualization GUI is an indispensable part of traffic simulation systems.
In MOSS, the simulator supports to output archived files for visualization and has implemented a simple visualization GUI based on Tkinter\footnote{\url{https://docs.python.org/3/library/tkinter.html}} as shown in Figure~\ref{fig:vis}(a), which can display the motion status of vehicles and the passability of lanes at different times.
On the online platform website of MOSS, the output results of the simulator will be stored in a database and presented through a WebGL-based visualization GUI powered by Deck.gl\footnote{\url{https://deck.gl/}} as shown in Figure~\ref{fig:vis}(b).
Due to the convenience of web visualization development, the web visualization currently provides more realistic vehicle models and presents more information.

\subsubsection{Python API}


Inspired by CityFlow~\cite{zhang2019cityflow}, we also implement C and Python bindings using pybind11, providing Python Application Programming Interfaces (APIs) in the form of a Python package for the convenience of researchers.
Compared to CityFlow, which mainly focuses on traffic signal optimization, we provide more query and modification APIs for lane speed limits, lane restrictions, and dynamic lanes to adapt to more refined traffic management optimization tasks besides traffic signal optimization.
We also provide query APIs for the travel time and distance of traffic participants, as well as various debugging information.
Additionally, to match the execution efficiency of the GPU-accelerated simulator, shared memory optimization with numpy is introduced in the simulator to avoid the overhead of copying during large-scale data reads.
Batch writing APIs are also provided to avoid the overhead of multiple interface calls.

\subsubsection{SUMO Converter}
To help SUMO users to quickly migrate to MOSS, we develop the SUMO Converter to convert SUMO format to MOSS format, which includes the following two components. 
\begin{itemize}
    \item \textbf{Map Converter}.
    The tool support conversion from SUMO road networks, including geometry positions and topology connections, etc., to MOSS format. 
    SUMO traffic light conversion is also supported.
\end{itemize}
\begin{itemize} 
    \item \textbf{Trip Converter}.
    The tool support to convert SUMO trip including routes, trips and flows to MOSS format.
\end{itemize}

\section{Evaluation}\label{sec:evaluate}

In the section, we performed three sets of experiments to answer the following research questions:
\begin{itemize}
    \item RQ1: What is the computational efficiency of MOSS in large-scale scenarios, and can it outperform the microscopic or even mesoscopic simulators?
    \item RQ2: Can MOSS simulate realistic results with real-world travel demand input?
    \item RQ3: What is the realism of the OD generation method of MOSS, compared to the baseline methods?
\end{itemize}

Moreover, we built two demonstrations to show MOSS's efforts in openness and easy-to-use, and support for large-scale traffic optimization, respectively.

\subsection{Computational Efficiency}

\begin{figure}[t]
  \centering
  \includegraphics[width=1\linewidth]{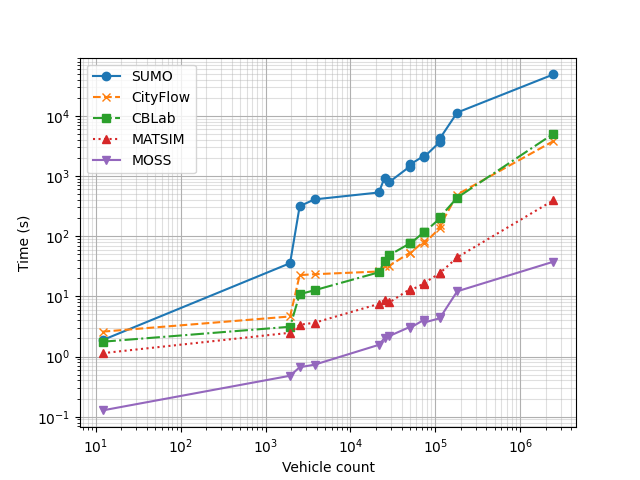}
  \caption{Comparison of simulation time consumption with the number of vehicles. (best viewed in color)}\label{fig:efficiency}
  \vspace{-0.5cm}
\end{figure}

To demonstrate the efficiency of MOSS, we conducted a series of experiments. 
We built 6 real-world cities' road networks and generated 9 road
networks with grid numbers ranging from 16 to 93564 as our test dataset.
The dataset includes different scale scenarios, covering from about ${10^0}$ to ${10^6}$ vehicles.
Regarding the baselines, we chose 3 widely used microscopic simulators including SUMO~\cite{behrisch2011sumo}, CityFlow~\cite{zhang2019cityflow}, and CBLab~\cite{liang2023cblab} and the popular mesoscopic simulator MATSim~\cite{w2016multi}.
All simulations were conducted in the same environment, utilizing an Intel(R) Xeon(R) Platinum 8462Y CPU and an NVIDIA GeForce RTX 4090 GPU. 
The baseline simulation system ran on the CPU, employing 32 threads for testing, while MOSS utilized 32 threads on the CPU and additionally used the GPU.
In our experiment, we simulated 3600 steps at a one-second interval for all datasets and performed three runs for each scenario.
The average of the three runs was reported as the final result.
The results shown in Figure~\ref{fig:efficiency} indicate that MOSS has a significant performance advantage over all baseline methods for all scales of scenarios.
On the largest dataset with 2,464,950 vehicles, the running time of MOSS is 37.70s, which is a 100.97 times improvement compared to the best microscopic simulator CityFlow (3806.7s) and a 10.49 times improvement compared to the mesoscopic simulator MATSim (395.48s).
The result fully illustrates the efficiency of MOSS, making model simplification in large-scale scenarios no longer necessary.

\subsection{Realism of the Simulator}

\begin{figure}[t]
  \centering
    \subfloat[MOSS simulation result]{\includegraphics[width=0.5\linewidth]{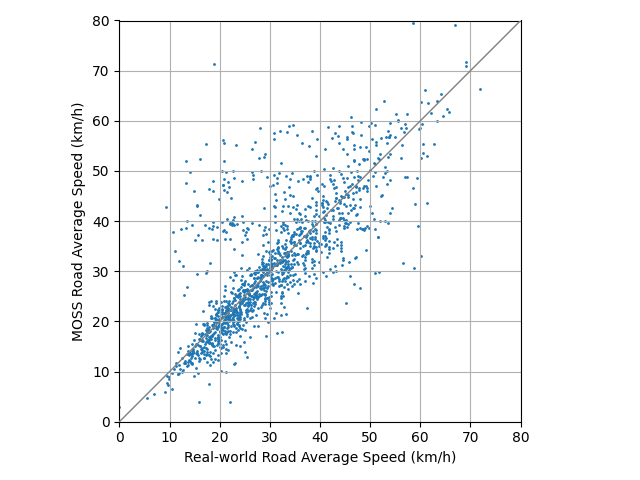}}
	\subfloat[CityFlow simulation result]{\includegraphics[width=0.5\linewidth]{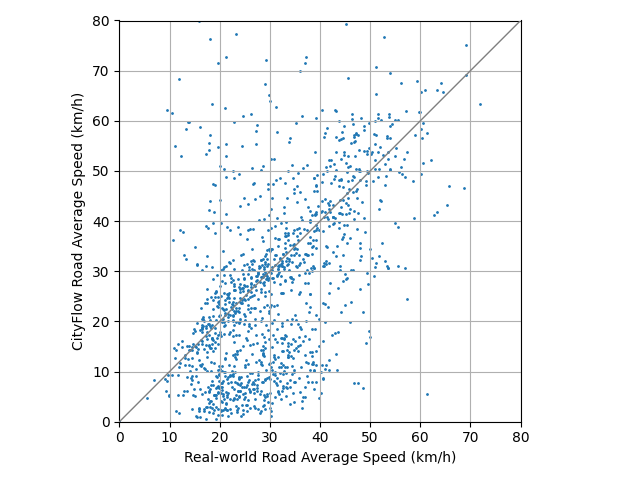}}
  \caption{Comparison of real-world and simulated average vehicle speeds.}\label{fig:speed}
  \vspace{-0.5cm}
\end{figure}
\begin{figure*}[t]
  \centering
  \includegraphics[width=0.8\linewidth]{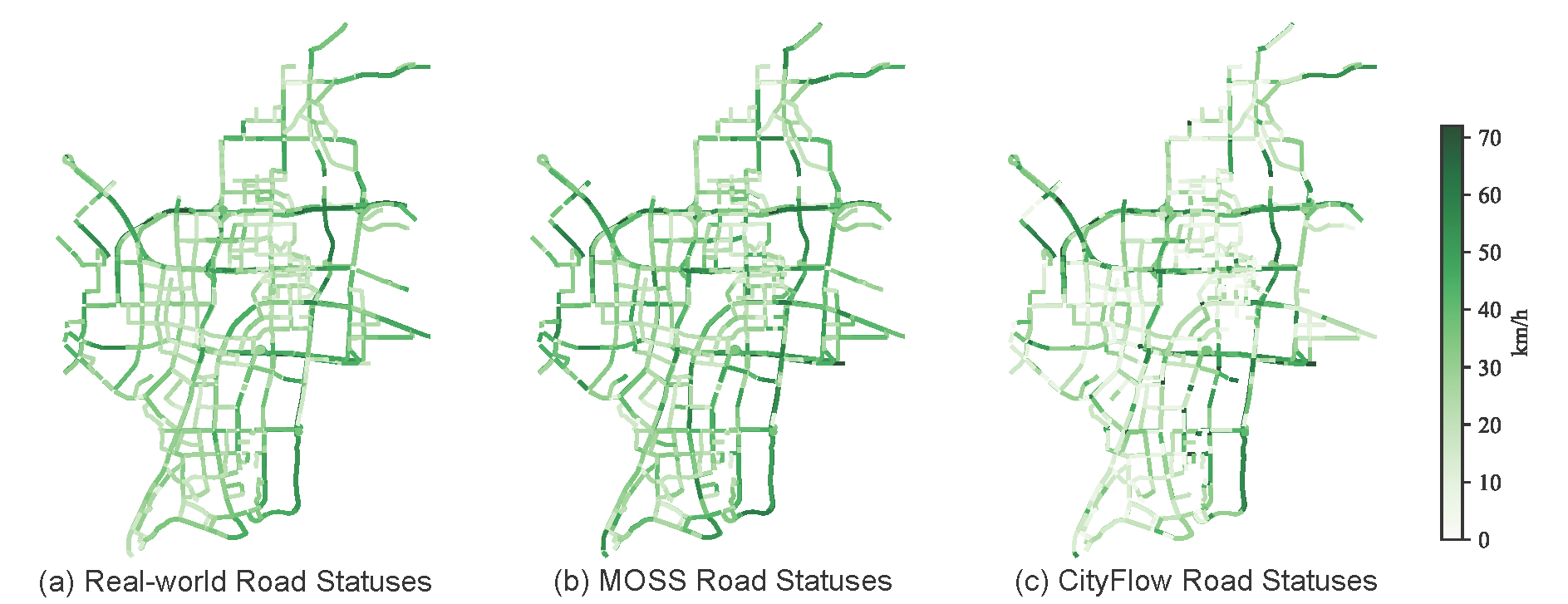}
  \caption{Visualization and comparison of road traffic status from (a) the dataset, (b) MOSS, and (c) CityFlow.
  (best viewed in color)}\label{fig:road_status}
  \vspace{-0.5cm}
\end{figure*}

To illustrate the ability of the MOSS simulator to accurately reflect real-world traffic status, the experiment inputs real-world travel demand data into the simulator and compares the similarity between the results and the real traffic situation data in order to assess its realism.
The real-world travel demand dataset is from Shenzhen, China~\cite{yu2023city}.
The dataset matches 4.22 million vehicle GPS trajectories with the road network to obtain the average speed for each road as the real-world traffic status.
It also uses images of vehicles captured by traffic cameras to identify the same vehicles and recover their trajectories as the real-world travel demand data.
The number of recovered trips within the Nanshan District of Shenzhen is 156,856.
The experiment compares the simulated average vehicle speeds of 1,341 roads from 8 to 9 AM, as simulated by MOSS and the best-performing baseline microscopic simulation system CityFlow.
The results are presented in Figure~\ref{fig:speed}, where the average road speeds obtained from the MOSS simulator are closer to the real-world data compared to CityFlow.
The RMSE of road speeds from MOSS has decreased from 16 km/h of CityFlow to 8.5 km/h, an improvement of 46.8\%, while the correlation coefficient has increased from 0.5286 to 0.7691, an improvement of 45.5\%.
We also visualize the spatial distribution of road statuses simulated by both simulators and compared them with real-world data in Figure~\ref{fig:road_status}.
It can be observed that the road statuses simulated by MOSS are closer to the actual ones.
In contrast, the CityFlow's results show lower speeds in some dense road network areas and create fake congestion.
The above results indicate that the MOSS simulator has a high degree of realism.

\subsection{Realism of OD Generation}\label{sec:exp:demand}

To illustrate the realism and global usability of the OD generation method in MOSS, we designed two sets of experiments.
The first set compares the performance of our method with baseline methods on the Longitudinal Employer-Household Dynamics
Origin-Destination Employment Statistics (LODES) collected by the National Census Bureau of the United States~\cite{lodes2024us}.
The second set employs our model trained on the LODES dataset to generate OD matrices in multiple cities worldwide and compares them with local real-world data.

\begin{figure}[t]
  \centering
    \subfloat[CPC$\uparrow$ of different models]{\includegraphics[width=0.5\linewidth]{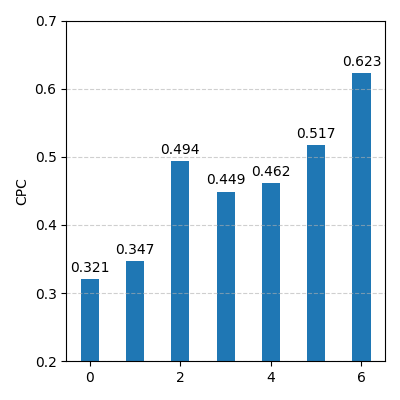}}
	\subfloat[RMSE$\downarrow$ of different models]{\includegraphics[width=0.5\linewidth]{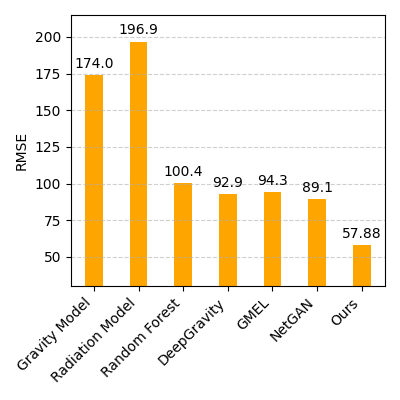}}
  \caption{The comparison of the OD matrix generation tested on the dataset from the United States.}\label{fig:odmetric}
  \vspace{-0.5cm}
\end{figure}

In the first set of experiments, the LODES dataset includes OD matrices for 2,275 counties in the United States, with the data divided into an 8:1:1 ratio for training, validation, and testing, respectively.
The baseline models to be compared include both rule-based and data-driven approaches.
Rule-based model contains gravity mode and radiation model
Data-driven models contains random forest~\cite{pourebrahim2019trip}, DeepGravity~\cite{simini2021deep}, GMEL~\cite{liu2020learning} and NetGAN~\cite{bojchevski2018netgan}.
The baseline models use traditional well-structured sociodemographic and geographical features as input for training, while our method utilizes publicly available satellite imagery as the training input.
The common part of commuting (CPC) and the root mean square error (RMSE) are chosen as the evaluation metrics following \cite{rong2023city}.
The results of this set of experiments are presented in Figure~\ref{fig:odmetric}.
The results indicate that our model, based on publicly available satellite imagery, can generate OD matrices that are more accurate than those generated by existing rule-based or data-driven baseline methods.
This suggests that our method has better realism in generating OD matrices for travel demand in traffic simulation.

\begin{table}[t]
    \centering
    \caption{Comparison of generated OD matrices with real ones for multiple cities or countries around the world.}
    \begin{tabular}{cc}
        \hline
        City/Country & Spearman correlation coefficient \\
        \hline
        Beijing, China & 0.741 \\
        Shanghai, China & 0.488 \\
        Chengdu, China & 0.417 \\
        Paris, French & 0.465 \\
        Sydney, Australia & 0.503 \\
        Rio de Janeiro, Brazil & 0.816 \\
        Senegal & 0.477 \\
        \hline
    \end{tabular}
    \label{tab:odcity}
    \vspace{-0.3cm}
\end{table}

\begin{figure}[t]
	\centering
	\subfloat[Real-world OD matrix]{\includegraphics[width=0.47\linewidth]{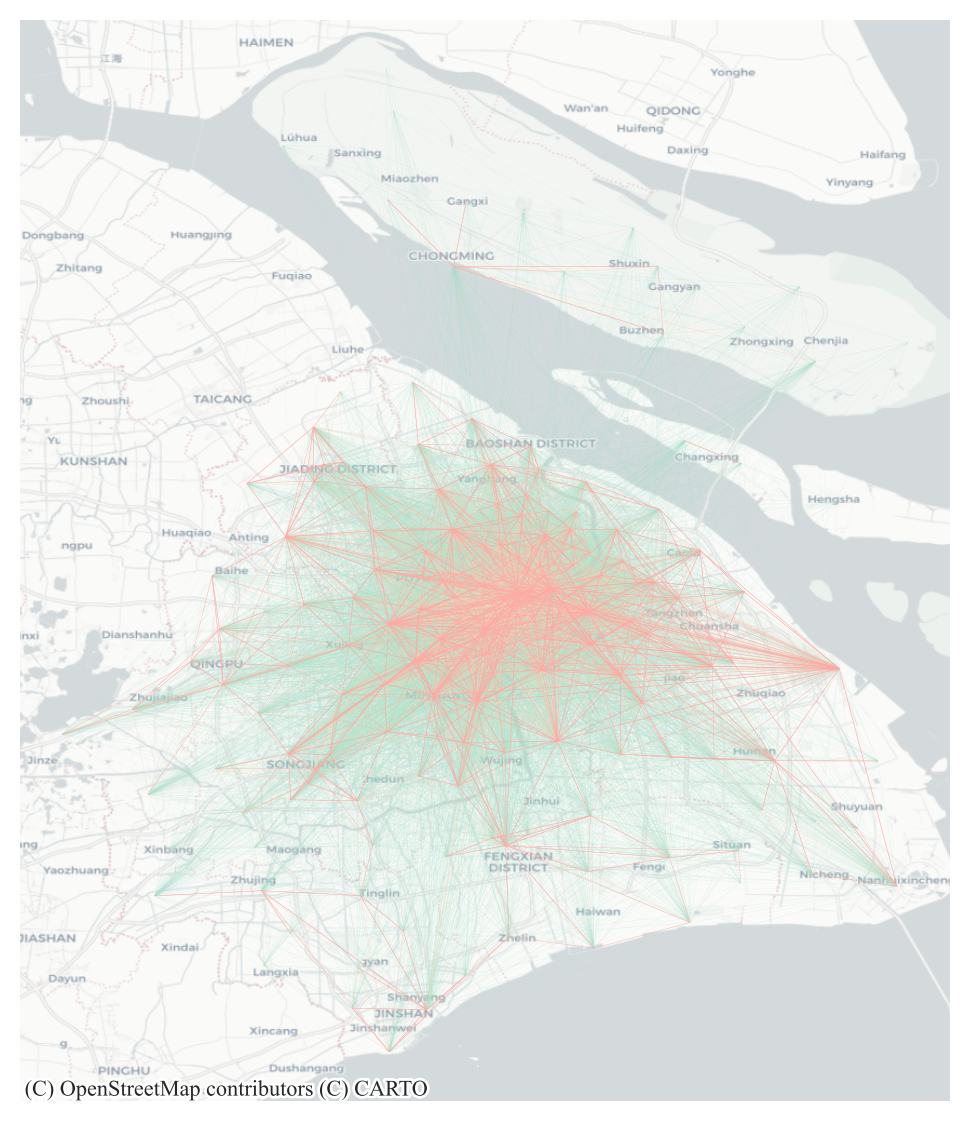}}
	\subfloat[Generated OD matrix]{\includegraphics[width=0.47\linewidth]{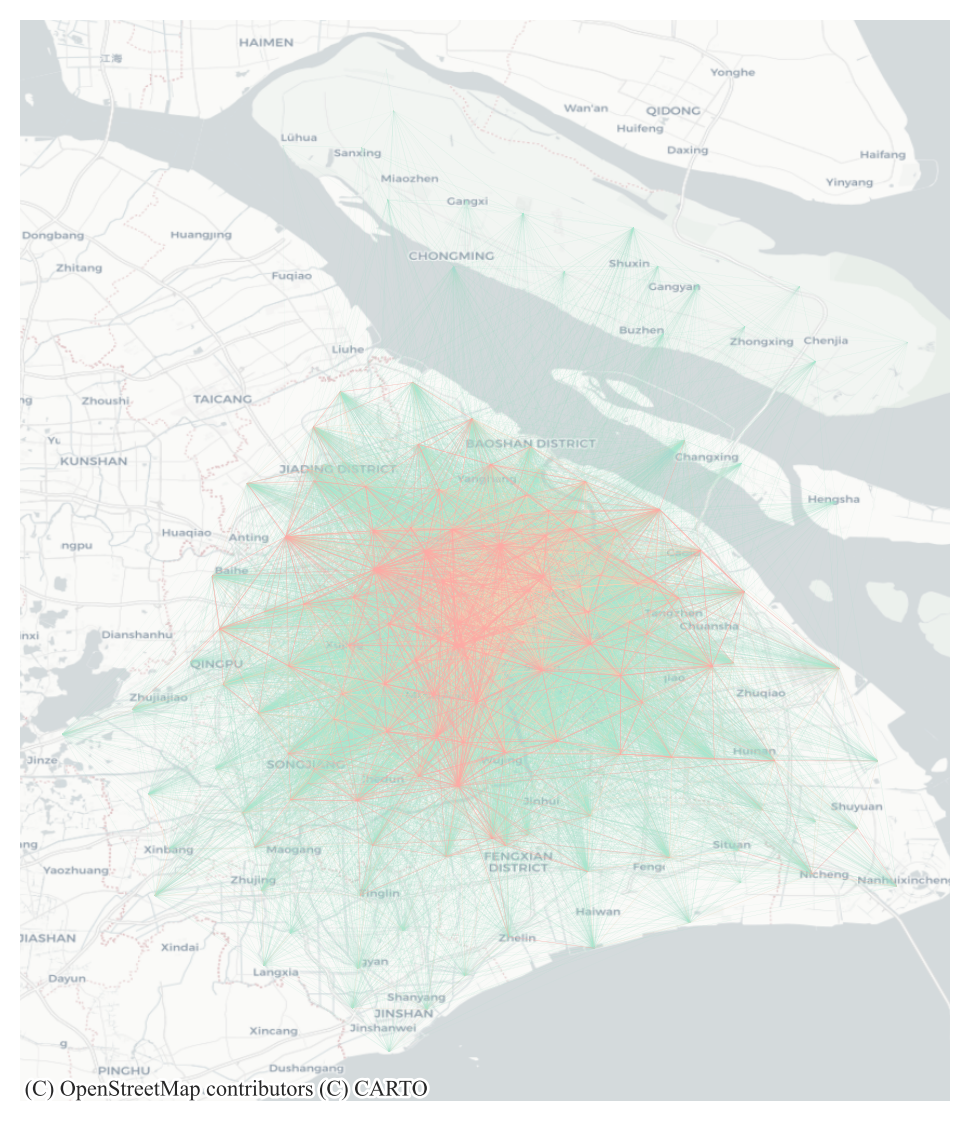}}
	\caption{Visualization of the real-world OD matrices in Shanghai with the ones generated by our proposed approach. (best viewed in color)}\label{fig:odcity}
    \vspace{-0.5cm}
\end{figure}

In the second set of experiments, we utilized the model trained on the LODES dataset and the relative satellite imagery to infer the OD matrices of typical cities or countries worldwide, assessing its global usability and realism.
We selected cities or countries from China, French, Australia, Brazil and Senegal as targets for generation.
Due to the difficulty in collecting real-world OD data, it is challenging to obtain ground truth across different countries and regions.
Therefore, we adopted various ancillary data from LBS providers, government census data,  call detail records, etc. that reflect the spatial distribution of human flows to perform correlation analysis and visualization with the generated ones to verify the realism of our generative method.
Table~\ref{tab:odcity} and Figure~\ref{fig:odcity} respectively present the correlation analysis results of cities worldwide using Spearman correlation coefficient and the visualization comparison between the generated results and actual data for Shanghai, China.
From these results, we can observe that the method of generating OD matrices based on publicly available satellite imagery is applicable in cities worldwide, capable of producing results with relatively high realism.
Therefore, the method can help researchers to obtain realistic OD matrices worldwide and improve the realism and usability of large-scale simulations.


\subsection{Zero Code Wizard}


To help researchers interested in MOSS quickly experience the system, we released a web-based Zero Code Wizard\footnote{\url{https://moss.fiblab.net/wizard}} based on the open-source MOSS simulator and MOSS Toolbox.
The Zero Code Wizard consists of the following four stages:
\begin{enumerate}
    \item \textbf{Road network construction.}
    After logging in, users can first select a preset city or choose any areas on the map to construct a new road network.
    The backend program will automatically obtain raw data from OpenStreetMap, first converting it to GeoJSON format, then constructing it into a Protobuf format map, saving it to the storage system, and binding it to the user.
    \item \textbf{Demand generation.}
    Subsequently, users can perform OD generation on the constructed map and select parameters for converting the OD matrix into individual travel demand, with the travel demand generation results also being saved.
    \item \textbf{Simulation.}
    In this stage, users can choose the simulation time range, the spatial range for saving microscopic simulation results, signal control algorithms, and parameters, etc.
    \item  \textbf{Visualization.}
    After completing the simulation, users can view the microscopic motion states of traffic participants and macroscopic statistical results obtained from the simulation in the online visualization GUI.
\end{enumerate}

With the help of the wizard, users can simulate the traffic of any city worldwide with only a few mouse clicks and parameter settings.
This significantly lowers the barrier to large-scale traffic simulation.


\subsection{Large-scale Transportation Optimization}

\begin{table}[t]
\caption{Average travel time (s) comparison among different traffic signal control strategies.}
\label{tab:att}
\centering
\begin{tabular}{cccc}
\hline
   \textbf{City}                  & \textbf{Shanghai} & \textbf{Hangzhou} & \textbf{Nanchang} \\
   \hline
\textbf{FP}          & 3814.72     & 3324.96     & 2764.90     \\
\textbf{MP}          & 3728.89     & 3244.23     & 2689.54     \\
\textbf{PPO}         & 3665.43     & 3148.40     & 2515.72     \\
\textbf{Improvement} & 1.70\%      & 2.95\%      & 6.46\%     \\
\hline
\end{tabular}
\vspace{-0.3cm}
\end{table}

To show that MOSS can support large-scale reinforcement learning applications, we selected the broadly concerned task of fine-grained traffic management optimization, i.e., traffic signal control optimization, as an example.
In the demonstration, we select three Chinese cities for large-scale signal control optimization, including Shanghai with 7,428 intersections, Hangzhou with 2,929 intersections, and Nanchang with 1,274 intersections.
The methods of comparison include the fixed-phase program (FP), the maximum pressure algorithm (MP)~\cite{varaiya2013max}, and a PPO-based~\cite{schulman2017proximal} reinforcement learning algorithm (PPO).
As shown in Table~\ref{tab:att}, through MOSS simulation and optimization with reinforcement learning algorithms, the learning-based strategy can produce certain improvements in traffic efficiency with the average travel time (ATT) metric on cities of different scales, indicating that MOSS can effectively support large-scale city-level traffic optimization.

The code of this demonstration is also open-source and can be accessed at \url{https://github.com/tsinghua-fib-lab/moss-opt-showcases}.

\medskip

In conclusion, the above experiments and demonstrations show that our proposed MOSS has reached a cutting-edge level in terms of efficiency, realism, and openness, and is able to effectively realize large-scale traffic simulation and support traffic optimization in large-scale scenarios.
\section{Conclusion}\label{sec:conclusion}

In the paper, we introduce MOSS, a large-scale open microscopic traffic simulation system.
We use GPU acceleration to provide massive computational power to solve the dilemma of scale and realism in traffic simulation, enabling large-scale microscopic traffic simulation.
We also address the realism and accessibility of large-scale travel demand data through publicly available data represented by satellite imagery and cutting-edge generative AI methods.
MOSS will aid in expanding the research and applications of ITS to larger road networks, a wider range of countries and regions, and the adoption of more advanced computer decision-making technologies to optimize urban transportation systems.
As future research, we are building reinforcement learning solutions for traffic management problems such as large-scale city-level intersection turn proportionality assignment, dynamic lanes, and congestion pricing with the support of MOSS.




\bibliographystyle{IEEEtran}
\bibliography{IEEEabrv,root}

\end{document}